\documentclass[manuscript]{aastex}

\def\gsim{\mathrel{\raise.5ex\hbox{$>$}\mkern-14mu
             \lower0.6ex\hbox{$\sim$}}}
\def\lsim{\mathrel{\raise.3ex\hbox{$<$}\mkern-14mu
             \lower0.6ex\hbox{$\sim$}}}
\def\simless{\mathbin{\lower 3pt\hbox
     {$\rlap{\raise 5pt\hbox{$\char'074$}}\mathchar"7218$}}}   
\def\simmore{\mathbin{\lower 3pt\hbox
     {$\rlap{\raise 5pt\hbox{$\char'076$}}\mathchar"7218$}}}   

\def\Sw{{\em Swift}}
\def\Ss{{\em Swift}~}
\def\Ep{$E_{\rm pk}$}
\def\Eps{$E_{\rm pk}$~}
\def\R{${\cal R}$}
\def\Rs{${\cal R}$~}

\shorttitle{The statistics of BAT-to-XRT flux ratio in GRB}
\shortauthors{Kazanas et al.}

\begin{document}

\title{The statistics of BAT-to-XRT flux ratio in GRB:
Evidence for a characteristic value and its implications}

\author{D. Kazanas\altaffilmark{1}, J. Racusin\altaffilmark{1}, J. Sultana\altaffilmark{2},
and A. Mastichiadis\altaffilmark{3}}

\email{Demos.Kazanas@nasa.gov}

\baselineskip= 14pt

\altaffiltext{1}{Astrophysics Science Division, NASA/Goddard Space
Flight Center, Greenbelt, MD 20771 USA.}
\altaffiltext{2}{Mathematics Department, Faculty of Science,
University of Malta, Msida MSD2080 Malta.}
\altaffiltext{3}{Department of Physics, University of Athens,
Panepistimiopolis, GR 15783, Zografos, Greece}

\begin{abstract}

We present the statistics of the ratio, \R, between the prompt and
afterglow ``plateau" fluxes of GRB. This we define as the ratio
between the mean prompt energy flux in {\em Swift} BAT and the {\em
Swift} XRT one, immediately following the steep transition between
these two states and the beginning of the afterglow stage referred
to as the ``plateau". Like the distribution of many other GRB
observables, the histogram of \Rs is log-normal with maximum at a
value ${\cal R}_m \simeq 2,000$, FWHM of about 2 decades and with
the entire distribution spanning about 5 decades in the value of \R.
We note that the peak of the distribution is close to the
proton-to-electron mass ratio $({\cal R}_m \simeq m_p/m_e = 1836)$,
as proposed to be the case in an earlier publication, on the basis
of a specific model of the GRB dissipation process. It therefore
appears that, in addition to the values of the energy of peak
luminosity ${E_{\rm pk}\sim m_{e} c^2}$, GRB present us with one
more quantity with an apparent characteristic value. The fact that
the values of both these quantities (\Ep~and \R) are consistent with
the same specific model invoked to account for the efficient
conversion of their relativistic proton energies to electrons,
argues favorably for its underlying assumptions.

\end{abstract}


\section{Introduction}
Gamma-Ray Bursts (GRB), $\gamma-$ray emission events at cosmological
distances by relativistically moving (of Lorentz factors $\Gamma
\sim 200$) plasmas,  remain enigmatic despite much observational and
theoretical progress over the past twenty five years. Furthermore
they present us with new puzzles each time instrumentation and
ensuing observations that measure any of their attributes improve
significantly. The most recent such puzzle is that of the shape of
their afterglow light curves as determined by \Sw, the subject of
the present note.

%

While the association of GRB with RBW is generally accepted, there
are still serious gaps in our fundamental understanding of their
underlying physics. For instance, the presence of the GRB
relativistic outflows with the necessary Lorentz factors requires
flows with energy-to-mass ratios larger than $\sim 100$, a condition
not naturally encountered in most astrophysical plasmas. Similarly,
bearing in mind that in a RBW the proton to electron postshock
pressure ratios are roughly in proportion to that of their mass,
i.e. $m_p/m_e \sim 2000$, one would expect a very small radiative
efficiency ($\eta \sim 1/ 2000$) for these shocks, considering that
protons are generally inefficient radiators. This issue is usually
resolved by {\em assuming} the efficient transfer of energy from
protons to electrons to keep the two components in rough pressure
equipartition, thus guaranteeing the efficient emission of radiation
by the synchrotron emitting electrons. Another open issue associated
with the GRB prompt stage is the rather limited range of the energy
of peak luminosity $E_{\rm pk} \sim 0.3$ MeV (extending on occasions
to a few MeV), intriguingly close to the electron rest mass on the
Earth frame, but not on the GRB rest frame, considering the large
Lorentz factors of their blast waves. This characteristic energy is
{\em the} defining attribute of the GRB prompt emission since
emission at this energy declines very fast GRB develop in time and
enters their, less variable, afterglow stage.

Since their discovery by {\em BeppoSAX}, GRB afterglows have been
considered a distinct and separate phase of the GRB phenomenon.
However, to the best of our knowledge, there is no formal criterion
for the transition from the prompt to the afterglow GRB phase; it is
generally considered that the prompt GRB emission continues until
the RBW has reached its deceleration radius $R_D$, the radius at
which the RBW has swept-up mass-energy $Mc^2 \simeq m_p c^2 nR_D^3
\simeq E/\Gamma^2$ ($E$ is total energy of the RBW, $n$ the
circumburst density - assumed to be constant - and $\Gamma$ its
asymptotic Lorentz factor). Beyond this radius, the GRB Lorentz
factor $\Gamma$ decreases gradually (e.g. $\Gamma \propto R^{-3/2}$
for adiabatic evolution in a uniform density medium) as the added
inertia of the swept-up medium exceeds the value that would permit
expansion at constant $\Gamma$. The afterglow emission, just as that
of the prompt phase, is considered to be synchrotron radiation by
shock accelerated electrons, in rough equipartition with the protons
of the postshock region. Under these assumptions, one can compute
for the GRB afterglow stage the resulting spectrum and its time
evolution. The computation of the GRB X-ray flux evolution in time
was first performed by \citet{sar98} for spherically symmetric
outflows, and by \citet{sar99} for jet-like outflows, assuming the
spectrum of electrons injected at the shock to be $dN/dE \propto
E^{-p}, ~ p \gsim 2$. The resulting X-ray light curves were, then,
shown to be power laws in time, $F_X \propto t^{-\alpha}, ~ \alpha
\gsim 1$, in broad agreement with the sparsely sampled afterglow
light curves of the pre--{\em Swift} era.

The launch of \Ss and its ability to follow closely the evolution of
GRBs from their prompt ($\gamma-$ray emitting) to the afterglow
(X--ray emitting) stages provided yet another set of unexpected,
puzzling facts, grossly inconsistent with expectations based on the
models described above \citep{tagliaferri05,nousek,zhang06,evans09}:
Instead of the predicted decrease of their X-ray flux as $L_X
\propto t^{-\alpha}, \alpha \gsim 1$, the decrease is much steeper
($\alpha \sim 3 - 6$), followed either ({\em i}$\,$) by a much
shallower section (referred to as the ``plateau") ($\alpha \sim 0$),
which is succeeded for $t > T_{\rm brk} \sim 10^3-10^5$ sec by a
decline of $\alpha \simeq 1$ or ({\em ii}$\,$) not by a plateau, but
by a more conventional decline ($\alpha \simeq 1$) (these light
curves exhibit also occasional large amplitude flares which we will
not discuss at present).

The commonly accepted account of the steep light curve decline
segment is that of high latitude late emission \citep[see][\S 3.1.1
and references therein]{Z07}, even though the time evolution of such
emission, while steeper ($\alpha \simeq 2$) than that of Sari et al.
(1998), it is still less steep than what is  observed and certainly
at odds with declines as steep as $t^{-6}$. {However, \citet{GG09}
suggested that this process can account for power laws steeper than 2
and applied to fit the steep decay light curves of a sample of 12
GRBs \citep{Willn09} as steep as $t^{-3.5}$. } An alternative
explanation attributes this to the form of the underlying electron
distribution function (Kazanas et al. 2007; Giannios \& Spitkovsky
2009), while Petropoulou et al. (2011), by adjusting the maximum
energy $\gamma_{max}$ of the electron distribution, interpret the
steep decline segment as synchrotron emission by the fast cooling,
high energy cutoff of the electron distribution function and the
``plateau" segment to inverse Compton by its more slowly varying low
energy section.

The plateau segment, because it follows that of the very steep flux
decline, gives the impression of a distinct and completely separate
emission from that of the GRB prompt phase. For this reason, it was
proposed (not unreasonably) to represent an additional injection of
energy by the GRB central source, separate from that producing the
shorter but brighter prompt emission \citep{WillOBr}. Along similar
lines, \citet{Gompertz14} even specify this additional injection as
due to the propeller effect of an underlying magnetar that powers
the entire burst. Anyway, these attempts to account for the behavior
of the GRB afterglow light curves were devised to model these
specific features, without any reference to or consideration of the
broader properties of the entire burst. Several treatments have
focused on just the properties of the plateau phase itself. Of these
we mention those of \citet{Lei11} and \citet{Matzner12} who, by
fitting the spectro-temporal evolution of the plateau segment of the
light curves of several GRB, conclude that this emission takes place
before the RBW has reached its deceleration radius $R_D$.

An altogether different approach to the afterglow evolution has been
that of \citet{SKM13}. This is different in that the afterglow
evolution, including all its details, is produced as an integral
part of the evolution of the entire burst, beginning with the
accelerating phase of the RBW  and continuing with its dissipation
and prompt emission, including also the correct value the energy of
the GRB prompt phase emission, \Ep. The central notion of this model
is a radiative instability that converts the relativistic proton
energy behind the RBW forward shock to $e^+e^-$--pairs through the
$p \, \gamma \rightarrow p \, e^+e^-$ reaction; the pairs then
produce more synchrotron photons, which produce more pairs and so on
\citep{kaz02,mas06,mas08,mas09}. This instability requires that the
column of relativistic protons in the postshock region be larger
than a critical value, in direct analogy with a supercritical
nuclear pile (hence the nomenclature of the model; put simply, one
cannot accumulate arbitrarily large columns of relativistic protons
for the same reason that one cannot accumulate arbitrarily large
amounts of U$^{235}$: they explode!). However, besides this
condition on the proton column, the instability imposes a kinematic
constraint on the synchrotron photon energy because the synchrotron
photons emitted by the $e^+e^-$--pairs must be able to pair-produce
in collisions with the protons. This effectively requires the RBW
Lorentz factor to be larger than a critical value {$\Gamma_c^5 \, b
\simeq 1$}, a demand imposed by the kinematic threshold of the above
reaction ($b = B/B_{\rm cr}$ is the postshock value of the GRB
magnetic field with $B_{\rm cr} = 4.4 \times 10^{13}$ G the value of
the critical magnetic field). Incidentally, the energy of peak
emission of the prompt phase of this model, {\em on the Earth
frame}, is also (in units of $m_e c^2$) \Ep$\simeq\Gamma^5 b \simeq
1$, so this process provides also a reason for the observed value of
\Eps of the prompt GRB phase.

One of the crucial elements of this model is the upstream scattering
of the RBW synchrotron radiation and its re-interception by it. As
noted in \citet{mas08,mas09} and more specifically in
\citet[][hereafter SKM13]{SKM13} this process induces a radiation
reaction on the RBW and causes a (relatively) small ($\sim
30\%-50\%)$ reduction of its Lorentz factor over a radius $\Delta R
\lsim R$. Even though small, this reduction pushes $\Gamma$ below
the threshold value $\Gamma_c$ and hence arrests the transfer of
energy from the RBW relativistic protons into $e^+e^-$--pairs. As
noted in {SKM13} this results in an abrupt reduction of the RBW
radiative flux (the steep decline) and the GRB enters its afterglow
stage. The reduction in flux should be by a factor roughly equal to
the proton-to-electron mass ratio, i.e. by $\sim m_p/m_e \simeq
2000$, since the emitted radiation now comes from the cooling of
{\em only the electrons} swept-up by the RBW. For the reasons
explained in the paragraph above, the RBW continues its expansion at
the new, constant, value of $\Gamma$ (until it reaches its
deceleration radius) by emitting radiation only by the newly
swept-up electrons. Because of the constancy of the Lorentz factor
in this stage, depending on the rate of decrease of the RGW magnetic
field with radius or the density profile of the ambient medium
\citep{Matzner12}, the ensuing synchrotron emission could be
constant, decreasing or even increasing with time. Finally, beyond
the deceleration radius, which occurs at time $T_{\rm brk}$, a more
conventional traditional decrease in its flux ensues, consistent
with the standard models.

Recently, an interpretation of the afterglow idiosyncrasies, similar
in some respects to that of {SKM13}, suggested by \citet{DMacF14},
who proposed that a ``top-heavy" jet would produce behavior similar
to that observed. The main point behind this approach is, in
essence, similar to that of our model (SKM13), namely a change in
the Lorentz factor of the RBW on a radial scale $\Delta R$ of order
or shorter than the local RBW radius, but much shorter than its
deceleration radius $R_D$. The difference of this model from that of
SKM13 is that this change in $\Gamma$ is achieved by adding inertia
(rather than negative momentum as in SKM13) onto the RBW over a
distance $\Delta R \lsim R$, {\em before} it has reached its
deceleration radius $R_D$. Just like in SKM13 the RBW achieves a new
Lorentz factor $\Gamma^{\prime} (< \Gamma)$ and expands at this
value until it piles-up enough inertia to continue its decrease at
the conventional rate. One should note here that, in distinction
with the model of SKM13, among other things, this model at its
present stage does not imply a specific value for the ratio ${\cal
R}$ of GRB luminosities between their prompt and plateau stages.

The specific value of the ratio  of the GRB fluxes between their
prompt phase and the afterglow plateau, implied by the model put
forward by {SKM13}, gave us the impetus to compile and present in
this paper the distribution of this ratio in a large number of GRB
afterglows of the \Sw-{\em XRT} repository \citep{evans09,evans10}.
This is presented in the next section. Along with this we also
present a correlation between $L_{\rm iso}$, {the peak isotropic
luminosity} of the prompt emission and the afterglow X-ray
luminosity $L_X$ at time $t = T_{\rm brk}$, the time the X-ray
afterglow resumes its more conventional {decay}. We finish in \S 3
with our conclusions and some discussion.

\section{The Prompt-to-Afterglow Flux Ratios}

Motivated by the arguments presented above, we have searched the \Ss
data base and compiled the ratios, \R, between the prompt and
afterglow GRB fluxes.
We have used fits to the average BAT light curves obtained from the
\Sw--{XRT} repository Burst Analyzer \citep{evans10} extrapolated
down to the 0.3--10 keV band from fits to the { XRT}, as described
in \citet{racusin09,racusin11}, to extract the flux at the
transition between steep decline and plateau in the afterglows
demonstrating that form.
In figures 1 and 2 we present three specific cases
of GRB transitions from the prompt to the afterglow stages with a
variety of post transition behaviors. The fast decline exponents
range between $\alpha \simeq -3$ and $\alpha \simeq -6$, while their
transition to $\alpha \simeq -1$ happens in all cases around $T_{\rm
brk} \simeq 10^4$ sec. The yellow arrows show the decrease in flux
between the {\em geometric mean} the highly variable prompt emission
and the end of the steep decline phase.

These figures indicate that these transitions have peculiarities of
their own. For example the decrease of GRB 050713A is by a factor
${\cal R} \simeq 10^3$, smaller than the $m_p/m_e$ ratio; however,
in the other two GRB the decrease as shown by the yellow arrows are
by factors $10^4$ and $10^6$, one of them significantly larger than
the $m_p/m_e$ ratio. On the other hand this last case, namely of GRB
120213A exhibits a rather peculiar two step transition with one step
decrease by a factor of $\sim 10^4$ followed by another one by a
factor of $10^2$, with the horizontal lines and the yellow arrow
indicating the fluxes considered by the employed algorithm. This
last case indicates that, besides applying a given algorithm, one
may have to scrutinize each such transition individually.

From the point of view of the data available in search of
correlations among the GRB attributes, the one proposed in SKM13 and
tested herein has a clear advantage in that it involves only flux
ratios rather than absolute values (whether luminosities, time lags
or values of \Ep) as is the case with many of the GRB produced
correlations (e.g. the Lag-Luminosity relation, the \Ep--$E_{\rm
iso}$,  the $L_X - T_{\rm brk}$. etc. correlations). As such,
knowledge of the GRB redshift is not necessary, a fact that allows
the compilation of a large number of bursts. The ratios of the
prompt to afterglow fluxes were compiled from the {\em Swift}--XRT
repository and spans the period between December 2004 and March
2014. The results of this compilation are give in figure 3a, where
we present a histogram of the prompt to afterglow flux ratios as
determined by our algorithm, along with the $m_p/m_e$ ratio given by
the dashed line of this figure.

\begin{figure}[t]
\begin{center}$
\begin{array}{cc}
\includegraphics[trim=0in 0in 0in 0in,keepaspectratio=false,
width=3.0in,angle=-0,clip=false]{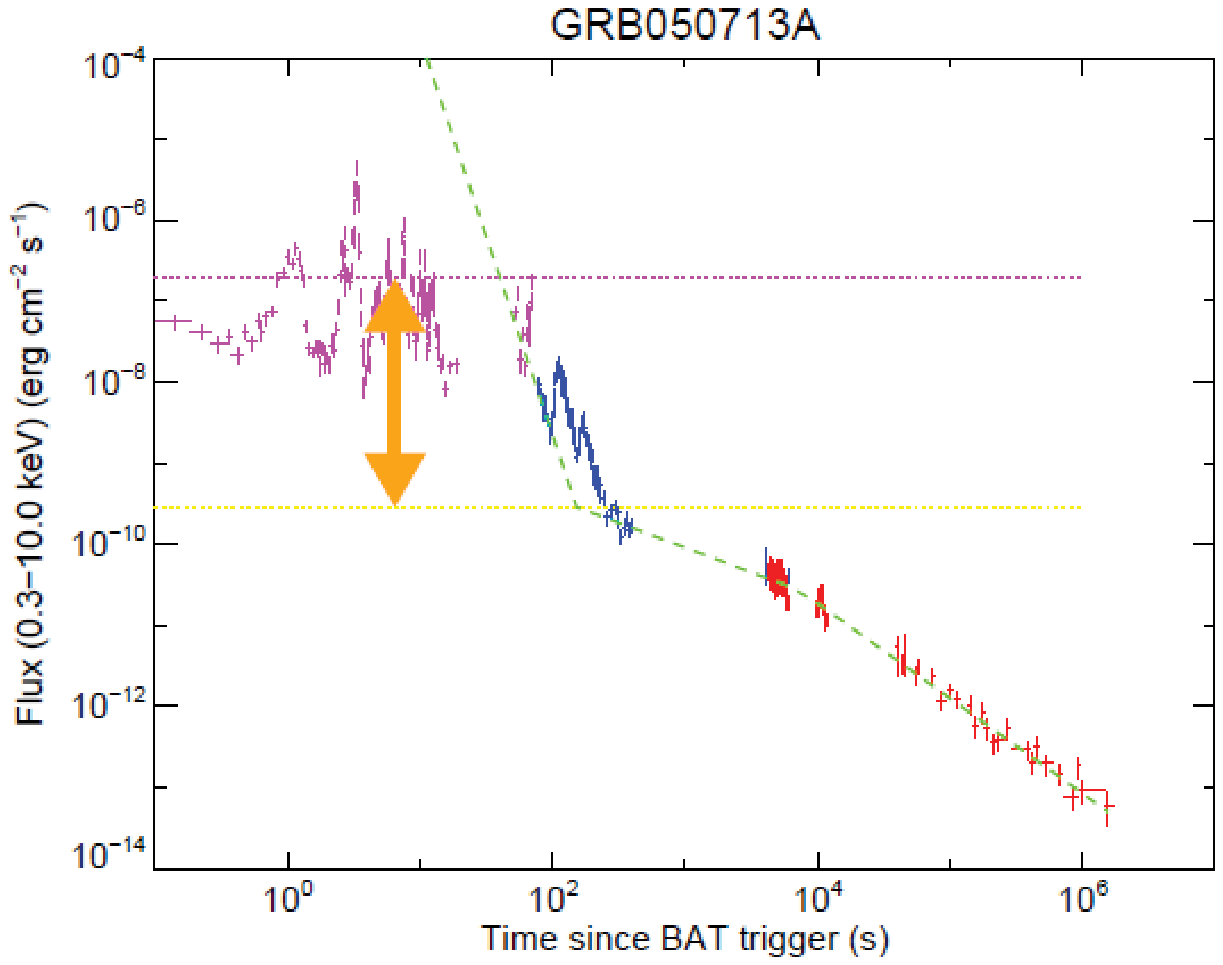}
&
\includegraphics[trim=0in 0in 0in 0in,keepaspectratio=false,
width=3.0in,angle=-0,clip=false]{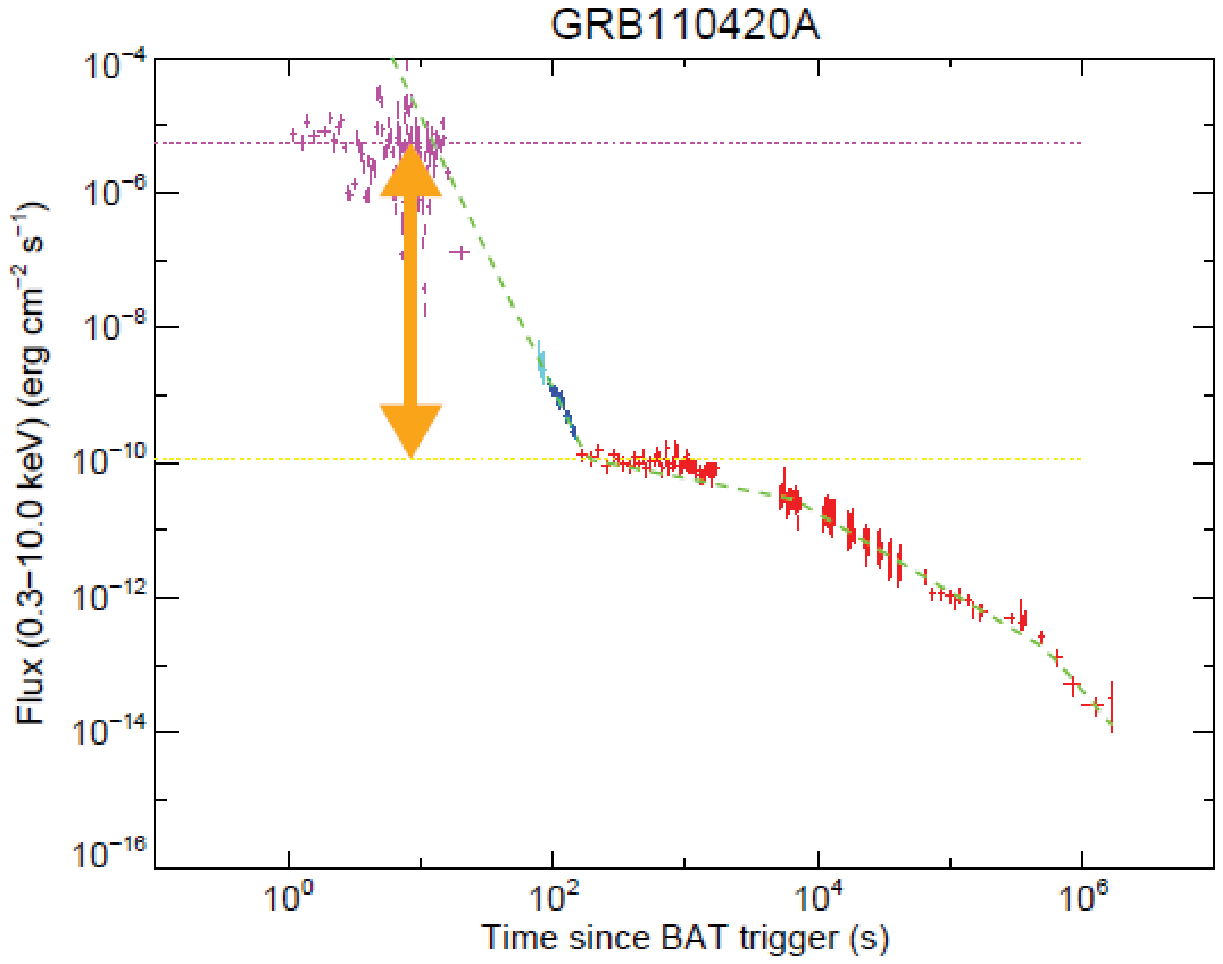}
\end{array}$
\end{center}
\vskip -20pt
\caption{\footnotesize  (a) The prompt to afterglow light curves of the gamma ray bursts
indicated on the figure. The prompt, transition and afterglow plateau stages are
apparent. The two dotted lines represent the mean prompt flux (top, purple line) and
afterglow (bottom, red line) fluxes involved in computing the flux ratios of these
two states. }
\label{fig:f1}
\end{figure}
%


The main result of our analysis is given in Fig. 3a where we present
a histogram of the logarithm of the BAT-to-XRT flux ratio, \R,
computed as described above along with a dashed vertical line that
indicates the value of the $m_p/m_e$ ratio. The distribution
exhibits a broad maximum at almost precisely the this value,
indicating the presence of a characteristic ratio between the prompt
and afterglow fluxes, as proposed in {SKM13}. The \R-distribution
appears to be log-normal, though its precise shape is not easy to
determine accurately. It spans 5-6 decades in \R, with a FWHM of
about 2 decades and a median value of $\log {\cal R}$ essentially
equal to that of $\log(m_p/m_e) \simeq 3.25$ and a slightly larger
medium value ($\simeq 10^4$) in sufficient agreement with the
suggestion of \citet{SKM13} to merit further consideration.

\begin{figure}[t]
\begin{center}$
\begin{array}{cc}
\includegraphics[trim=0in 0in 0in 0in,keepaspectratio=false,width=3.0in,
angle=-0,clip=false]{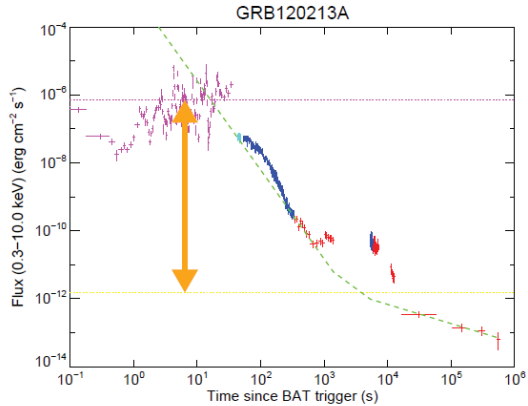}
\end{array}$
\end{center}
\vskip -20pt
\caption{\footnotesize  (a) The prompt to afterglow light curves of the gamma ray bursts
indicated on the figure. The prompt, transition and afterglow plateau stages are
apparent. The two dotted lines represent the mean prompt flux (top, purple line) and
afterglow (bottom, red line) fluxes involved in computing the flux ratios of these
two states. }
\label{fig:f2}
\end{figure}

As discussed above, there have been several searches for
correlations between GRB attributes involving either their prompt or
their afterglow emission. Amongst those of the first type we mention
the Lag--Luminosity relation \citep{nor00,nor02,S07} and the maximum
prompt emission (isotropic) luminosity $L_{\rm iso}$ and the peak
energy of the Band function $E_{\rm p}$ \citep{S07,WQD11}. Of the
correlations involving the GRB afterglow properties we mention the
relation between the X-ray luminosity $L_{X}$ at the end of the
plateau phase, and the rest-frame plateau-end time $T_{\rm brk}$,
beyond which the afterglow resumes the standard decline
\citep{dai08,dai10,dai13}.

In \citet{SK12} we combined these relations to show a significant
correlation between the timing properties of the prompt and
afterglow emissions, suggesting that these two phases are intimately
connected, despite the absence of the continuity between these two
phases implied by the models \citep{sar98}: In particular we showed
that the Lag$-$Luminosity relation of the prompt emission
extrapolates into the $L_X-T_{\rm brk}$ relation of their afterglow.
Now, this seems to be a little strange, since the time scale
associated with the Lag of the prompt emission, even though a time
scale, it is of a different character than the duration of the
afterglow plateau emission $T_{\rm brk}$. With the above discussion
and motivated by the histogram of Fig. 3a, we bypass the time
coordinate of the relation given in \citet{SK12} and plot in Fig. 3b
the maximum prompt isotropic luminosity $L_{\rm iso}$ vs. the X-ray
luminosity of the afterglow plateau segment at the time $T_{\rm
brk}$. There appears to be a correlation between these quantities. A
least squares fit gives the following relation between $L_{\rm iso}$
and $L_X$
\begin{equation}
\log L_{\rm iso} = (4.04 \pm 0.10) + (1.04 \pm 0.02) \log L_X
\end{equation}
with correlation coefficient $\rho = 0.69$. The ratio of these two
quantities appears consistent with that shown in Fig. 3a. Given the
difference in the choice of these samples and the slightly different
properties they depict, they appear to be consistent with each other
and the general premise of the prompt to afterglow luminosity
ratios.

\section{Discussion, Conclusions}

Motivated by the considerations put forward in \cite{SKM13} based on
the tenets of the SPM model of GRB dissipation, we have compiled the
flux ratios between the prompt and the afterglow plateau phases of
GRB. The important point to bear in mind is that a characteristic
value for the ratio \R, namely $m_p/m_e$ was given in SKM13 {\em
before} the compilation of the histogram of Fig. 3a; as such, this
constitutes a {\em prediction} of the model, one of the very few in
the GRB field of study.  A similar relation has also been found
between slightly different quantities of these two GRB phases, shown
in Fig. 3b, one that requires however knowledge of their redshifts.

%
\begin{figure}[t]
\begin{center}$
\begin{array}{cc}
\includegraphics[trim=0in 0in 0in
0in,keepaspectratio=false,width=3.3in,angle=-0,clip=false]
{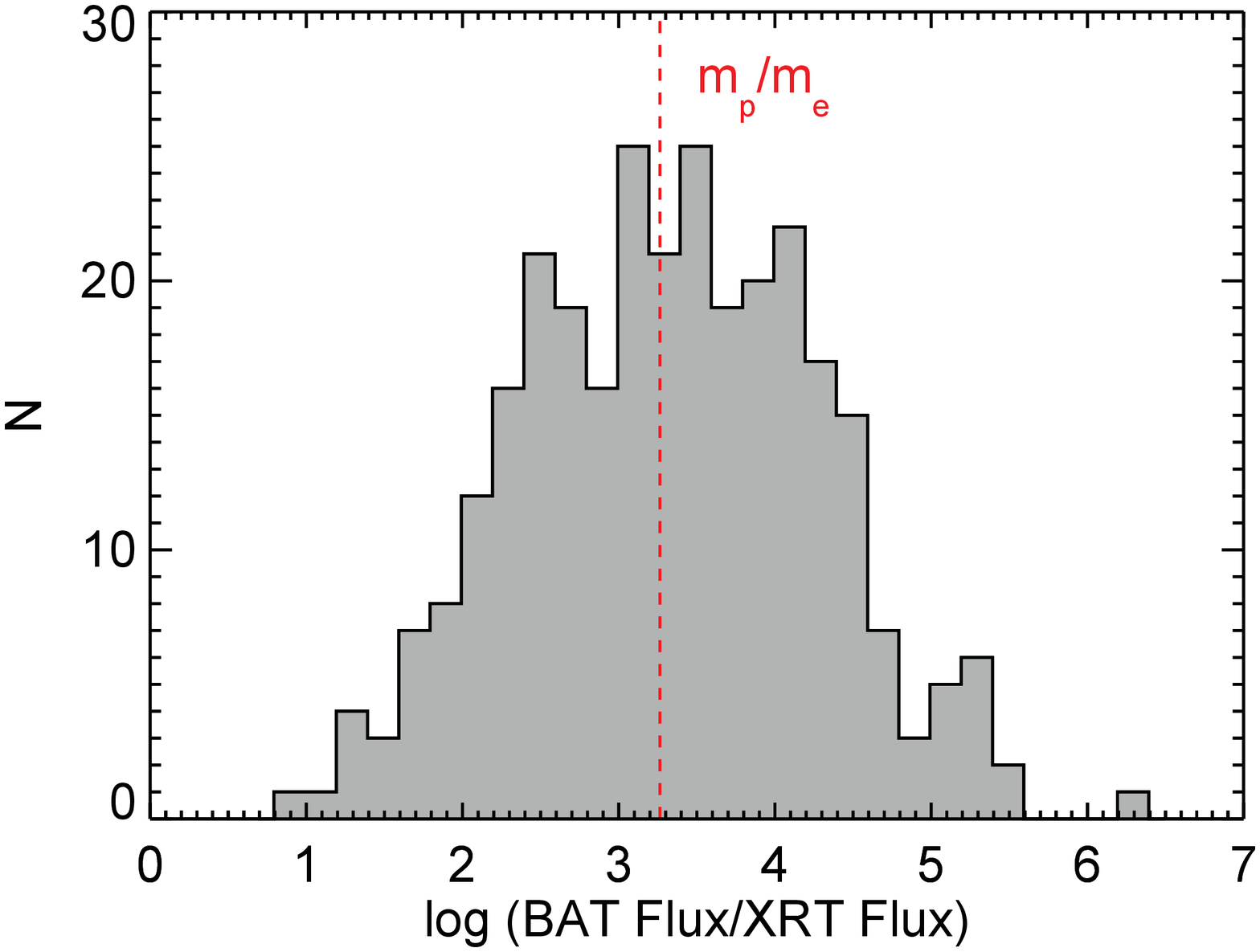} &
\includegraphics[trim=0in 0in 0in
0in,keepaspectratio=false,width=3.1in,angle=-0,clip=false]
{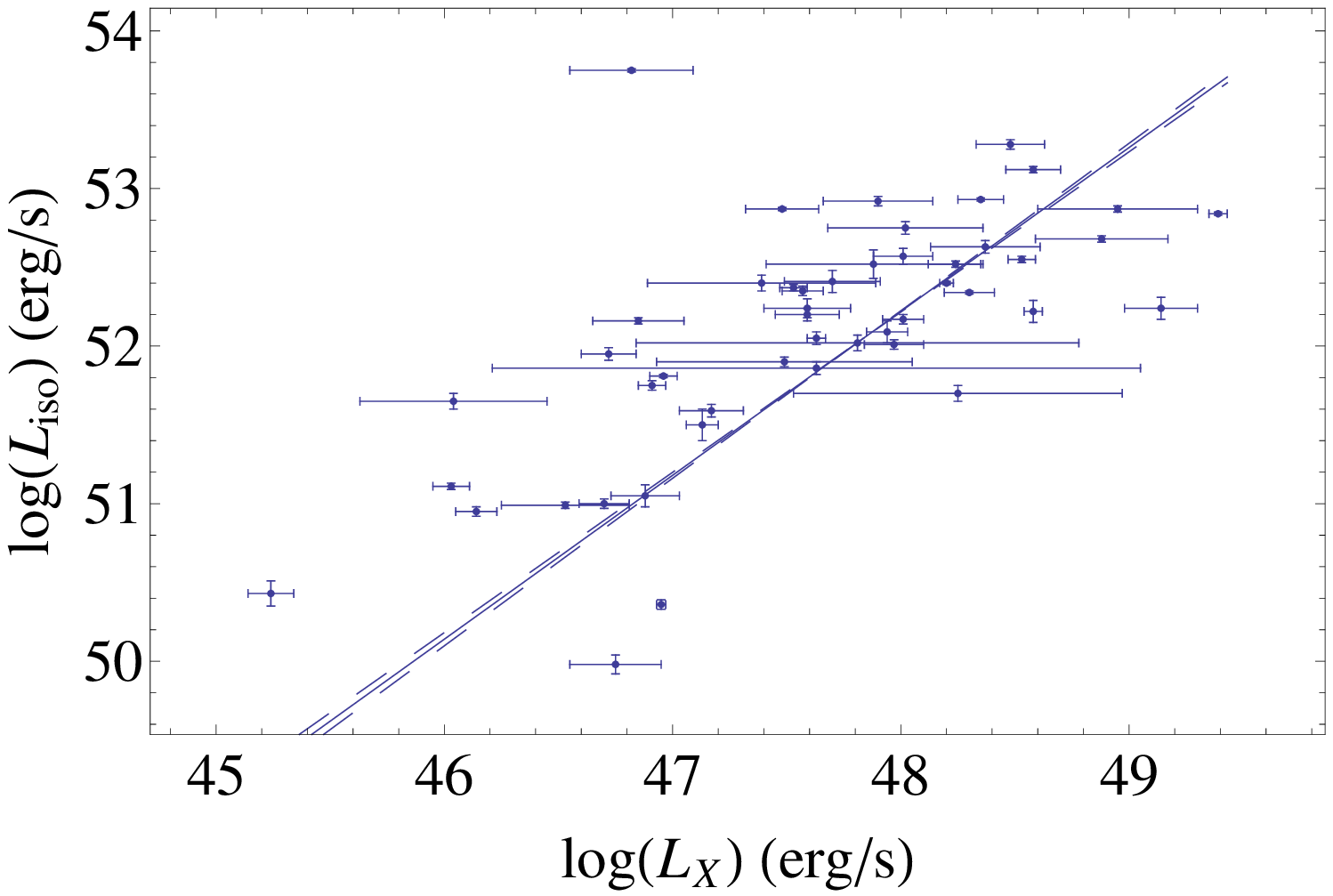}
\end{array}$
\end{center}
\vskip -20pt
\caption{\footnotesize  The histogram of the BAT to XRT flux ratio for a number of
{\em Swift} GRB. The distribution shows clearly a preferred value for this ratio of order
$\sim 10^3 - 10^4$. The vertical line shows also the proton to electron mass ratio $m_p/m_e$. }
\label{fig:f3}
\end{figure}

One must note at this point that though there is a maximum in the
distribution of flux ratios near the value $m_p/m_e$, the histogram
and correlation of Fig. 3 have a finite width. Thus there are bursts
with \Rs values as large as $10^6$ and as low as $10^2$. Figure 2
shows a burst with a particularly large value of this ratio. As
argued earlier, one could assign to this burst a value smaller than
that given by our algorithm, given the peculiar form of its
afterglow. On the other hand, if one takes into account that the
apparent luminosity of a relativistically moving source can have a
dependence on its Lorentz factor $\Gamma$ as strong as $\Gamma^4$,
even a small reduction in $\Gamma$ could increase the pre-to-post
prompt emission fluxes to values larger than $m_p/m_e$. Values of
${\cal R} < m_p/m_e$ seem to be more problematic. One possibility,
put forward in SKM13, is that not all protons ``are burnt" in prompt
phase, thus reducing the flux of this stage. {A related possibility
is that besides the postshock Maxwellian proton distribution, of
characteristic energy $\Gamma \, m_pc^2$, there is an additional,
non-thermal, power law proton population; because this population
extends to energies much higher than $\Gamma \, m_pc^2$, these
protons continue to fulfill the pair production condition and convert
their energy into pairs, as discussed in \cite{kaz07}, leading to a
reduced value for \R. The bright bursts GRB 110731A, GRB 130427A with
\R$\simeq 1000$ may in fact represent such cases (to keep the number
of free parameters to a minimum, our earlier treatments of the SPM
refrained from invoking non-thermal populations - a feature invoked
at will and expediently in all GRB models; this does not mean that
they are necessarily absent, however we prefer to invoke them only as
a last resort). } Finally, it is possible  that the angle between the
edge of the jet to the observer's line of sight, $\theta$, is
slightly larger than $1/\Gamma$, yielding a reduced relativistic
boosting for the prompt emission. After the RBW slow-down, the
smaller value of $\Gamma$ allows the observer's line of sight to
``peer" directly into the relativistic outflow, thereby reducing the
ratio of the pre-to-post prompt emission fluxes. {Independent of the
details of reason for which ${\cal R} < m_p/m_e$, the existence of
this characteristic value in the \R--distribution, provides a new
selection criterion by which we can distinguish the GRB properties
(e.g. Lags, \Ep, $L_{\rm iso}$, $E_{\rm iso},$ etc.) to get possibly
novel clues into the physics of GRB emission.}  We hope to return to
this issue in a future publication.

We conclude by pointing out that in our work, as in that of
\citet{DMacF14}, the GRB afterglow plateau is associated with the
dynamics of the RBW propagation rather than with the dynamics of the
GRB ``central engine", as is more commonly accepted; this fact
should give thrust to a direction of GRB modeling orthogonal to that
of heretofore. In support of the SPM is the fact that the
characteristic value of the prompt to afterglow flux ratio has been
predicted and the fact that this model provides a broader account of
the dissipation and spectral formation in GRB. In this respect, the
prompt and afterglow plateau stages are closely related, as pointed
out, among others, by \citet{SKM13}, based on the timing
correlations between these two phases.


\acknowledgments  J.S. gratefully acknowledges financial support
from the University of Malta during his visit at NASA-GSFC and the
hospitality of the Astrophysics Science Division of GSFC. D.K.
acknowledges support by {\em Swift} and {\em Fermi} GO grants. J.R.
acknowledges support by the {\em Swift} and {\em Fermi} projects.

\end{document}